\documentclass[aps,prl,superscriptaddress,amsfonts,amsmath,amssymb,showpacs,floatfix,reprint,longbibliography,footinbib]{revtex4-2}

\usepackage{url}
\usepackage{bm}
\usepackage{graphicx}
\usepackage{amsmath}
\usepackage{amstext}
\usepackage{amssymb}
\usepackage{amsfonts}
\usepackage{amsbsy}
\usepackage{verbatim}
\usepackage{color}
\usepackage[colorlinks=true, urlcolor=blue, linkcolor=blue, citecolor=blue, pdftex]{hyperref}
\usepackage[capitalise]{cleveref}
\usepackage{multirow}
\usepackage{floatrow}
\usepackage{float}
\usepackage{tikz}
\usepackage{gensymb}
\usepackage{textcomp}
\usepackage{enumitem}
\usepackage[version=3]{mhchem}
\usepackage{afterpage}
\usepackage{pbox}
\usepackage{makecell}
\usepackage{dsfont}
\usepackage{siunitx}
\usepackage{fancyhdr}
\usepackage{stackengine,wasysym,scalerel}
\usepackage{latexsym}
\usepackage{cancel}
\usepackage{array, multirow, bigdelim, makecell, booktabs}
\usepackage[misc]{ifsym}
\usepackage{times}
\usepackage[normalem]{ulem}

\begin{document}

\title{Candidate quantum disordered intermediate phase in the Heisenberg antiferromagnet\\ on the maple-leaf lattice}
  
\author{Lasse Gresista}
\email{gresista@thp.uni-koeln.de}
\affiliation{Institute for Theoretical Physics, University of Cologne, 50937 Cologne, Germany}
\affiliation{Department of Physics and Quantum Centre of Excellence for Diamond and Emergent Materials (QuCenDiEM), Indian Institute of Technology Madras, Chennai 600036, India}
\author{Ciar\'an Hickey}
\affiliation{Institute for Theoretical Physics, University of Cologne, 50937 Cologne, Germany}
\affiliation{School of Physics, University College Dublin, Belfield, Dublin 4, Ireland}
\affiliation{Centre for Quantum Engineering, Science, and Technology, University College Dublin, Dublin 4, Ireland}
\author{Simon Trebst}
\affiliation{Institute for Theoretical Physics, University of Cologne, 50937 Cologne, Germany}
\author{Yasir Iqbal}
\email{yiqbal@physics.iitm.ac.in}
\affiliation{Department of Physics and Quantum Centre of Excellence for Diamond and Emergent Materials (QuCenDiEM), Indian Institute of Technology Madras, Chennai 600036, India}
  \date{\today}

\begin{abstract}
Quantum antiferromagnets on geometrically frustrated lattices have long attracted interest for the formation
of quantum disordered states and the possible emergence of quantum spin liquid (QSL) ground states.
Here we turn to the nearest-neighbor spin-$1/2$ Heisenberg antiferromagnet on the maple-leaf lattice, which
is known to relieve frustration by the formation of canted $120^{\circ}$ magnetic order or valence bond crystal order
when varying the bond anisotropy. 
Employing a pseudo-fermion functional renormalization group approach to assess its ground state phase diagram
in detail, we present evidence for a QSL regime sandwiched between these two limiting phases.
The formation of such a QSL might signal proximity to a possible deconfined quantum critical point from which it emerges, 
and that is potentially accessible by tuning the exchange couplings. 
Our conclusions are based on large-scale simulations involving a careful finite-size scaling analysis of the behavior 
of magnetic susceptibility and spin-spin correlation functions under renormalization group flow. 
\end{abstract}

\maketitle

{\it Introduction.} The search for quantum spin liquid (QSL)~\cite{Savary-2017} ground states in spin models on two-dimensional geometrically frustrated lattices is a highly pursued endeavor. The spin $S=1/2$ Heisenberg model with antiferromagnetic couplings on the kagome~\cite{Iqbal-2013,He-2017} and triangular~\cite{Iqbal-2016_tri,Hu-2019_dmrg} lattices serve as classic examples where QSL behavior is well-established. Here, a QSL emerges when quantum order-by-disorder fails to select a unique ground state out of a degenerate manifold typically located at transition points between classical magnetic orders. Much less explored are potential QSLs in models with proximate antiferromagnetic and valence bond crystal (VBC) orders. Such an occurrence of an intermediate QSL phase could potentially signal vicinity to a deconfined quantum critical point (DQCP)~\cite{Senthil-2004a,Senthil-2004b} out of which the QSL develops, with the DQCP approachable by tuning exchange couplings. This scenario is currently being debated for the $S=1/2$ Heisenberg antiferromagnet on the Shastry-Sutherland lattice~\cite{Lee-2019} where a QSL phase occupying a narrow region between the plaquette singlet and antiferromagnetic orders has been reported in Ref.~\cite{Yang-2022}. It has also recently been proposed that a gapless QSL develops out of a DQCP in the $S=1/2$ $J_{1}$\textendash$J_{2}$\textendash$J_{3}$ Heisenberg model on the square lattice~\cite{Liu-2022,Liu-2023}. 

For the $S=1/2$ Heisenberg antiferromagnet on the maple-leaf lattice (MLL), it has been reported in Ref.~\cite{ghosh2022} that, upon tuning the nearest-neighbor bond anisotropy, one can traverse a phase diagram comprising canted $120^\circ$ (c$120^\circ$) antiferromagnetic~\cite{Farnell-2011} and VBC orders \textemdash with the latter shown to be an exact dimer ground state (a product state of singlets on isolated bonds) beyond a critical value of bond anisotropy~\cite{ghosh2022}. However, it remains to be established whether the transition between these two phases is continuous via a DQCP, first-order, or whether an intermediate QSL can develop gradually emerging out of an underlying DQCP which is proximate in parameter space. In this manuscript, we probe for the existence of an intermediate QSL by mapping the quantum phase diagram via large scale state-of-the-art pseudo-fermion functional renormalization group (pf-FRG) calculations~\cite{Reuther-2010,muller2023pseudofermion} combined with systematic finite-size and finite RG scale analysis of the behavior of magnetic susceptibility and spin correlation functions. We present evidence for a QSL ground state sandwiched between the c$120\degree$ magnetic and VBC orders. Although we don't provide any  direct evidence for a proximate DQCP, the noticeable similarity to the Shastry Sutherland case~\cite{Lee-2019} may suggest an analogous transition, setting the stage for further investigations in this direction.

\begin{figure}[b]
    \centering
    \includegraphics{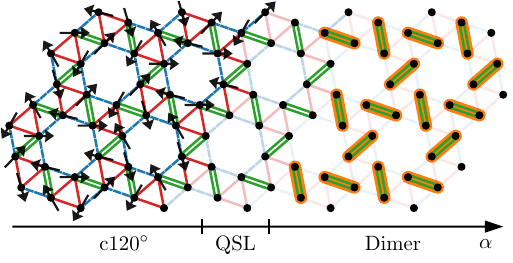}
    \caption{{\bf The maple-leaf lattice} admits three inequivalent nearest-neighbor bonds colored green, red and blue. Increasing the coupling anisotropy $\alpha$ on the green bonds induces a transition from a magnetically ordered phase (left) to a dimer phase (right), with an intermediate quantum disordered, putative spin liquid phase in between. The ordered phase exhibits canted $120^\circ$ order (c$120^\circ$), where spins on red triangles show local $120^\circ$ order, while spins on neighboring triangles are canted by an $\alpha$-dependent pitch angle $\Phi(\alpha)$.}
    \label{fig:lattice}
\end{figure}

\begin{figure*}
    \centering
    \includegraphics{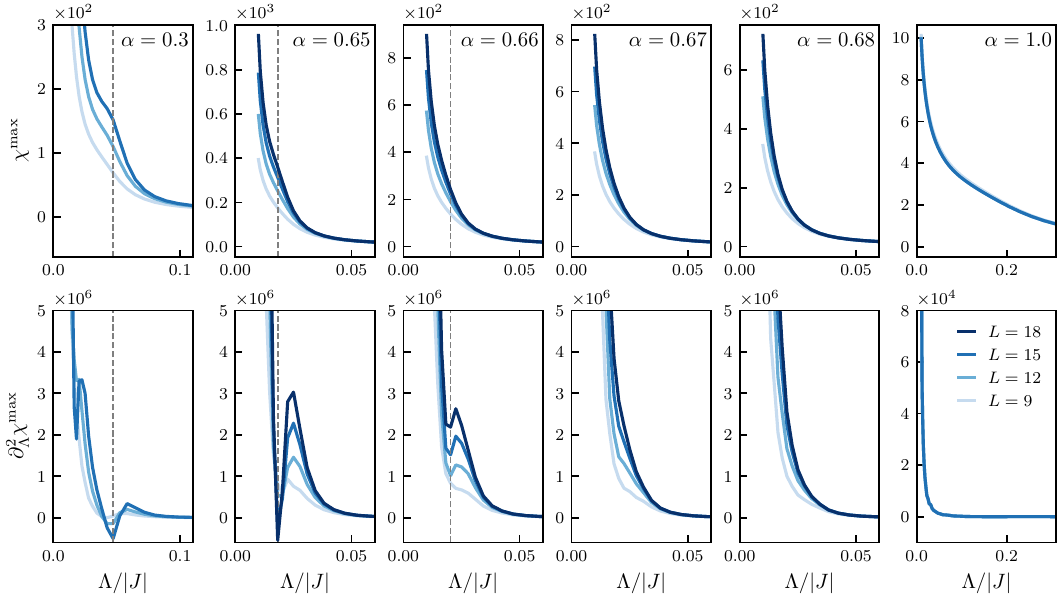}
    \caption{{\bf Renormalization group flow of the structure factor} at maximal intensity $\chi^\mathrm{max}$ (top) and its second derivative $\partial^2_\Lambda \chi^\mathrm{max}$ (bottom), at different values of $\alpha$. We identify a flow breakdown, indicated by the dashed gray lines, by a clear, lattice-size dependent non-monotonicity in the second derivative. }
    \label{fig:flows}
\end{figure*}

{\it Model.} The MLL [see Fig.~\ref{fig:lattice}] has a coordination number of $z=5$ and, by virtue of translation-rotation invariance between any pair of nearest-neighbor sites, is a uniform tiling~\cite{Betts-1995}. It belongs to the family of two-dimensional (2D) Archimedean lattices (i.e., with all lattice sites being symmetry equivalent) constructed by a periodic tessellation of regular polygons\textemdash triangles and hexagons. It can be obtained by a $1/7$ site-depletion of the triangular lattice. This results in a loss of reflection symmetry about any straight line through the lattice, and thus the point group of the MLL features only six-fold rotational symmetry about the center of hexagons, i.e., it has a $p6$ (No. 168) space group. Experimental realizations of the MLL are known to occur both in natural minerals~\cite{Fennell-2011,Kampf-2013,mills-2014} as well as synthetic crystals~\cite{Cave-2006,Aliev-2012,Haraguchi-2018,Haraguchi-2021}. It has a six-site crystallographic unit cell, and there exist three symmetry inequivalent nearest-neighbor bonds, which we color green, blue and red, as shown in Fig.~\ref{fig:lattice}. A generic nearest-neighbor Heisenberg model on the MLL may therefore be defined as
\begin{equation}
\label{eqn:heisenberg}
    \hat{\mathcal{H}} = 
    \sum_{\langle ij\rangle_G} J_G \hat{\mathbf{S}}_i \cdot \hat{\mathbf{S}}_j +
    \sum_{\langle ij\rangle_R} J_R \hat{\mathbf{S}}_i \cdot \hat{\mathbf{S}}_j +
    \sum_{\langle ij\rangle_B} J_B \hat{\mathbf{S}}_i \cdot \hat{\mathbf{S}}_j \,,
\end{equation}
where $\hat{\mathbf{S}}_i$ are $S=1/2$ operators at site $i$, $\langle ij \rangle_{G}$ denotes a sum over all green ($G$) nearest-neighbor bonds, and $J_{G}$ is the corresponding coupling. The same holds for red ($R$) and blue ($B$) bonds. Here, we consider the special case $J_R = J_B$, and only investigate the effect of a coupling anisotropy $\alpha$ defined via
\begin{equation}
\label{eqn:alpha}
 J_G = 2\alpha J_R = 2\alpha J_B.
\end{equation}
As already shown in Ref.~\cite{ghosh2022}, in this case the model is exactly solvable for $\alpha > \alpha_{b1} = 1.0$, where it exhibits an exact dimer ground state\textemdash a tensor product state of singlets covering the green bonds. Density matrix renormalization group (DMRG) calculations suggest that this state is stable down to $\alpha > \alpha_c^\mathrm{DMRG} \approx 0.675$. Additionally, below $\alpha < \alpha_{b2}\approx0.46$, it was shown that a magnetically ordered ground state exhibiting a canted $120^\circ$ order (c120$^\circ$) provides a better energy bound than the exact dimer singlet state. More importantly, since the DMRG analysis of Ref.~\cite{ghosh2022} only estimated the critical value of $\alpha$ where the dimer state terminates, the possibility of the system subsequently entering a more exotic quantum paramagnetic ground state, such as a quantum spin liquid, for $\alpha_c^\mathrm{DMRG} \lesssim 0.675$ remains unexplored.

{\it Methods and results.} To probe for the possible existence of such a quantum paramagnetic regime, we analyze model~\eqref{eqn:heisenberg} employing the pseudo-fermion functional renormalization group (pf-FRG) approach at zero temperature~\cite{muller2023pseudofermion}. The pf-FRG is, by now, an established method to distinguish between long-range magnetically ordered and quantum paramagnetic phases, and has additionally been used to determine dimerization or nematic tendencies in paramagnetic regimes lacking conventional dipolar magnetic order~\cite{Reuther-2010,Iqbal-2016_3d,Iqbal-2016_nem,Iqbal-2019_prx,Iqbal-2019_bcc,Astrakhantsev-2021,keles2022,Hagymasi-2022,Kiese-2023}. The main idea of the the pf-FRG is to introduce an infrared cutoff $\Lambda$, also called \emph{RG scale}, such that all correlation functions in the limit $\Lambda \to \infty$ are purely determined by the bare couplings in the spin Hamiltonian, and the physical correlation functions of the quantum theory are recovered for $\Lambda = 0$. The interpolation from high to low RG scales is described by a hierarchy of coupled integro-differential equations called \emph{flow equations}. 
These can, under certain approximations, be integrated numerically, starting from the large $\Lambda$ limit, which yields the flow of the two- and four-point correlation functions. We perform the integration by implementing the MLL into the \texttt{PFFRGSolver} Julia package
\cite{PFFRGSolver}, featuring state-of-the-art, adaptive integration schemes for the pf-FRG flow equations \cite{kiese2022}. In the integration, four-point correlations are parametrized by $n_\Omega = 35$ bosonic, $n_\nu = 40 \times 40$ fermionic Matsubara frequencies. We use lattice truncations of up to $L = 18$, i.e., correlations are set to zero beyond a bond distance of 18. This necessitates the solution of roughly $5.5 \cdot 10^8$ coupled differential equations, which we are able to reduce by a factor of six by utilizing the point group symmetry of the MLL.

\begin{figure}
    \centering
    \includegraphics{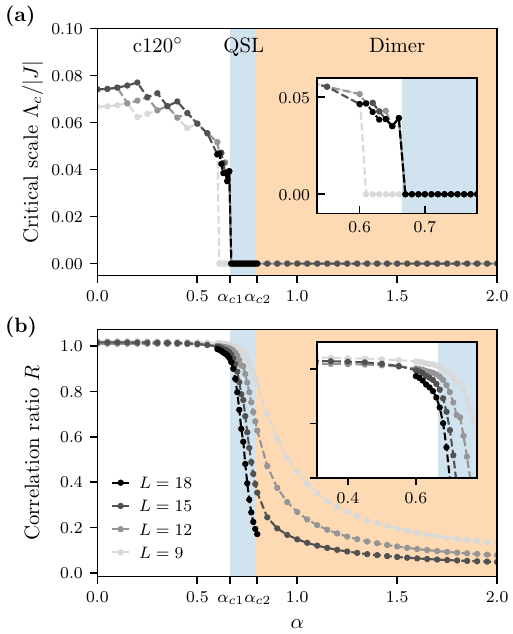}
    \caption{{\bf Melting of magnetic order}. \textbf{(a)} Critical scale  $\Lambda_c$ at which the pf-FRG flow shows an instability, for different lattice truncation lengths $L$. Below $\alpha < \alpha_{c1} \approx 0.66$ finite values of $\Lambda_c$ imply the formation of c$120^\circ$ magnetic order. Above $\alpha > \alpha_{c1}$, the absence of a flow breakdown ($\Lambda_c = 0$) implies a quantum paramagnetic ground state, where for $\alpha > \alpha_{c2} \approx 0.80$, the nature of correlations implies a dimer phase. Between the c$120^\circ$ and dimer phase at $\alpha_{c1} < \alpha < \alpha_{c2}$ a putative QSL phase emerges. \textbf{(b)} Correlation ratio $R$ as defined in Eq.~\eqref{eq:correlation-ratio} calculated at $\Lambda = 0.015/|J|$. The first decrease of $R$ coincides reasonably well with $\alpha_{c1}$, supporting the existence of a paramagnetic ground state for $\alpha > \alpha_{c1}$.}
    \label{fig:critical-scale}
\end{figure}

\begin{figure}
    \centering
    \includegraphics{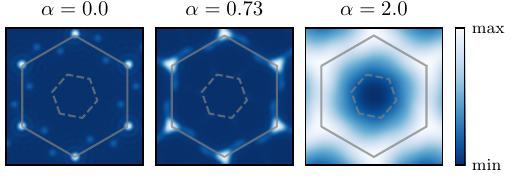}
    \caption{{\bf Static structure factor} in the c$120^\circ\ (\alpha = 0.0)$, QSL ($\alpha = 0.73$) and dimer ($\alpha = 2.0$) phase~\cite{Ghosh-2023_blue}. The dashed (solid) grey line indicates the first (extended) Brillouin zone of the MLL.}
    \label{fig:sfs}
\end{figure}

In order to differentiate between magnetically ordered and paramagnetic states, we calculate the flow of the (static) spin-spin correlations 
\begin{equation}
    \label{eq:correlations}
    \chi_{ij}^\Lambda = \int_0^\infty\!\!\! d\tau e^{i\omega \tau}\! \left\langle \hat{T}_\tau \hat{S}^z_i(\tau)\hat{S}^z_j(0)\right\rangle\Big|^\Lambda_{\omega = 0}
\end{equation}
from the two- and four-point correlations, where $\hat{T}_{\tau}$ is the time-ordering operator in imaginary time $\tau$ (due to the spin rotational symmetry of the Hamiltonian, the $xx$, $yy$ and $zz$-correlations are equivalent so it suffices to study just one of them). A straight-forward Fourier transformation of $\chi_{ij}^\Lambda$ yields the flow of the magnetic structure factor, i.e., static susceptibility. Onset of magnetic order manifests itself in the divergence of the structure factor flow (evaluated at the Bragg peak wave vectors of the incipient order) at some finite $\emph{critical scale}$ $\Lambda_c$\textemdash in 2D this divergence is an artifact of the truncation of the flow equations in contrast to 3D where it reflects a phase transition. Due to finite numerical resolution, however, the divergence -- or \emph{flow breakdown} -- is often considerably softened to a cusp or a kink, which sharpens with increasing lattice truncations $L$. The structure factor flow of a paramagnetic state, in contrast, is expected to be smooth and convex down to the smallest RG scales, independent of the lattice truncation. We consequently identify any non-monotonicity in the second derivative of the flow $\partial^2_\Lambda\chi$ that becomes more pronounced with increasing lattice truncation as indicative of a flow breakdown. Examples of the corresponding structure factor flows and their second derivatives are shown in Fig.~\ref{fig:flows} (with $|J|^2 = J_G^2 + J_R^2 + J_B^2$ as an energy scale for normalization). Using this criterion, we calculate the evolution of the critical scale $\Lambda_c$ as a function of $\alpha$. As shown in Fig.~\ref{fig:critical-scale}(a), for $\alpha < \alpha_{c1} \approx 0.66$ we find a magnetically ordered c120$^\circ$ phase. In this state, the $120^\circ$-ordered triangles spiral with an $\alpha$-dependent pitch, reminiscent of block-spiral magnetism~\cite{Herbrych-2020}. For $\alpha > \alpha_{c1}$, the flow shows no instability indicating a quantum paramagnetic phase. This is supported by a broadening of the maxima in the structure factor when entering the paramagnetic regime, as shown in Fig.~\ref{fig:sfs}. 

To quantify this effect, we calculate the correlation ratio $R$ \cite{kaul2015, pujari2016, yusuke2021} defined as
\begin{equation} \label{eq:correlation-ratio}
    R = 1 - \chi(\mathbf{K} + \boldsymbol{\delta})/\chi(\mathbf{K}),
\end{equation}
where $\mathbf{K}$ is the K-point of the extended Brillouin zone and $\boldsymbol{\delta}= 2\pi/\tilde{L} \ (1/\sqrt{3}, -1)^\top$ is the first reciprocal lattice vector of the triangular lattice underlying the MLL, scaled with the maximally allowed correlation length $\tilde{L}$ (in terms of real-space distance) for a given bond truncation length $L$. The result is shown in Fig.~\ref{fig:critical-scale}(b). The correlation ratio tends to $R = 1$ in the ordered phase, as the Bragg peak becomes increasingly sharp in the thermodynamic limit, and to $R \to 0$ in the disordered phase. The first decrease of $R$ coincides reasonably well with the critical $\alpha_{c1}$ determined by the flow breakdown analysis. We do not observe a clear crossing point for curves of different $L$, as might be expected at a phase transition \cite{kaul2015, pujari2016, yusuke2021}. We note, however, that the correlation ratio is calculated for constant $\Lambda/|J| = 0.015$, which lies below the flow breakdown in the ordered phase ($\alpha < \alpha_{c1}$) for which the numerical data may not be reliable, possibly impeding the formation of a clear crossing point.

Our estimate of $\alpha_{c1} \approx 0.66$ places the isotropic model [$\alpha=0.5$ in Eq.~\eqref{eqn:alpha}, i.e., $J_{G}=J_{R}=J_{B}$ in Eq.~\eqref{eqn:heisenberg}] inside the magnetically ordered regime with the spin configuration being a translationally invariant coplanar $120^\circ$ order with a six-sublattice structure~\cite{Schulenburg-2000}. Previous studies based on exact diagonalization, coupled cluster approximation, spin-wave and variational approaches have highlighted the precarious nature of the isotropic point finding either a quantum paramagnet or a weakly ordered state~\cite{Schulenburg-2000,Schmalfuss-2002,Farnell-2011,Farnell-2014,Farnell-2018,Makuta-2021}. This is, in some sense, reflective of the fact that among nonbipartite lattices, the coordination number $z=5$ of the MLL lies in between that of the triangular lattice ($z=6$), which has $120^\degree$ magnetic order~\cite{Capriotti-1999}, and the kagome lattice ($z=4$), which has a quantum paramagnetic ground state for the $S=1/2$ Heisenberg antiferromagnetic model.

\begin{figure}
    \centering
    \includegraphics{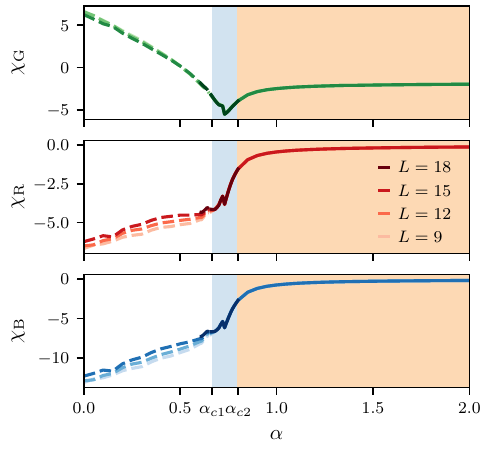}
    \caption{{\bf Nearest-neighbor spin-spin correlations} on green $\chi_G$, red $\chi_R$ and blue $\chi_B$ bonds [defined in Fig.~\ref{fig:lattice}] at $\Lambda/|J| = 0.01$ for different lattice truncations $L$. Above $\alpha_{c2} \approx 0.8$, the correlations $\chi_G$ tend towards a constant value while $\chi_R$ and $\chi_B$ tend to zero, indicating a dimer ground state covering the green bonds. Correlations only show a lattice size dependence in the long-range ordered c120$^\circ$ state (dashed lines), where the RG flow shows a breakdown at finite $\Lambda_c$.}
    \label{fig:bond-correlations}
\end{figure}

\begin{figure}
    \centering
    \includegraphics{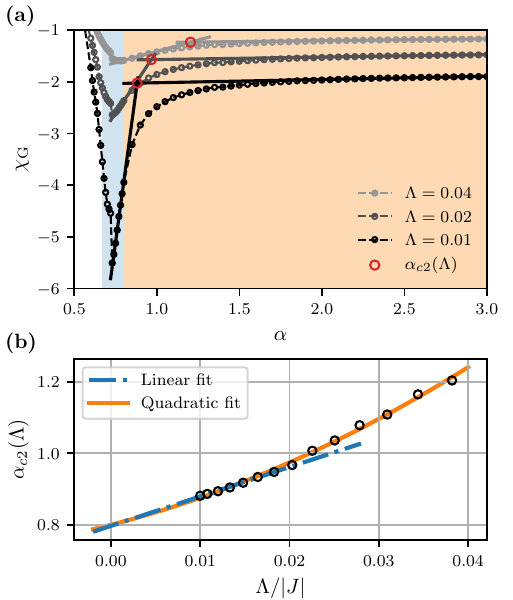}
    \caption{{\bf Transition into the dimer ground state.} \textbf{(a)} Spin-spin correlations on the green bonds $\chi_G$ for different RG scales $\Lambda$. In the dimer state the correlations tend to a constant value. The transition point into the dimer state is determined by the intersection of two linear regressions (solid black lines) for each $\Lambda$. Extrapolating the resulting $\alpha_{c2}(\Lambda)$ to $\Lambda = 0$, as shown in $\mathbf{(b)}$, leads to a critical value of $\alpha_{c2} \approx 0.8$. The data presented corresponds to $L = 15$, but the simulation outcomes are largely independent of the lattice truncation length within the disordered regime [see 
    Fig.~\ref{fig:bond-correlations}].}
    \label{fig:dimer-transition}
\end{figure}

To further characterize the nature of the putative quantum paramagnetic phase on the MLL, we calculate the nearest-neighbor spin-spin correlations on the green, red and blue bonds $\chi_{G/R/B}$ in the small $\Lambda$ limit [see Fig.~\ref{fig:bond-correlations}]. For large values of $\alpha$, the correlations on the green bonds $\chi_G$ rise sharply in a linear fashion and saturate to a constant value, while $\chi_B$ and $\chi_R$ rapidly go to zero. This RG flow fixed point indicates the onset of a robust phase with constant spin correlations on the green bonds. This indicates a progressively enhanced propensity towards the formation of a dimer-product state~\cite{Farnell-2011} where all the green bonds host a dimer singlet~\cite{Misguich-1999}, and which ultimately evolves to being the exact ground state~\cite{ghosh2022}. Such a behavior of the spin-spin correlation functions is akin to what has previously been observed in a pf-FRG analysis of the Shastry-Sutherland model~\cite{keles2022} which is also host to an exact dimer ground state. Additionally, the momentum resolved structure factor shows no sharp peaks compared to the magnetically ordered phase [see Fig.~\ref{fig:sfs}]. The rise of $\chi_G$ to its saturation value becomes more rapid when decreasing $\Lambda$. Following Ref.~\cite{keles2022}, we assume that, in the thermodynamic limit, on crossing the phase transition into the dimer phase the ground state immediately transitions into a product state, and is thus accompanied by an immediate saturation of the spin-spin correlations on the green bonds. In that case, we can determine the transition point $\alpha_{c2}$ to the dimer phase, as a function of $\Lambda$, by the intersection of two linear regressions [black lines in Fig.~\ref{fig:dimer-transition}(a)] and then extrapolate the result to $\Lambda = 0$ obtaining an estimate of $\alpha_{c2} \approx 0.80$, as is illustrated in Fig.~\ref{fig:dimer-transition}(b). This analysis implies that our estimate of the dimer crystal to be the ground state is for $\alpha > \alpha_{c2}$, slightly higher than the estimate of DMRG calculations \cite{ghosh2022}, which\textemdash as a variational approach targeting low-entanglement states \cite{Schollwoeck2011}\textemdash might have a tendency to overestimate the stability of the dimer crystal phase. Conversely, for $\alpha_{c1} \lesssim \alpha \lesssim \alpha_{c2}$, we find a quantum paramagnetic ground state (inferred from an absence of flow breakdown) whose correlations resemble that of the c120$^\circ$ state. Indeed, as one passes from the c120$^\circ$ state to the intermediate paramagnetic state and even into the dimer state, the structure factor features soft maxima at the location of the Bragg peaks of the c120$^\circ$ magnetically ordered state, which smoothly soften as $\alpha$ increases.

{\it Discussion.} We have presented evidence pointing to a QSL phase in the $S=1/2$ Heisenberg antiferromagnet on the MLL. Given that the QSL occupies an intermediate region in parameter space separating the c$120^{\circ}$ antiferromagnetic and VBC orders, further investigation is warranted concerning its possible origin from a proximate DQCP which can be accessed by tuning exchange couplings. Indeed, such an emergence of a gapless QSL from a DQCP has lately been discussed in the context of the $S=1/2$ $J_{1}$\textendash$J_{2}$\textendash$J_{3}$ Heisenberg model on the square lattice~\cite{Liu-2022}. The MLL model could potentially serve as a second rare example of a frustrated model host to both a QSL and DQCP, with the former originating from the latter, thereby highlighting their intrinsically intertwined nature. This would involve a mapping of the global phase diagram upon varying all symmetry inequivalent nearest-neighbor couplings and/or possibly including longer-range ones to establish whether the QSL terminates at a point and beyond which there is a direct c$120^\circ$-VBC transition. To this end, it would be worthwhile to employ other state-of-the-art numerical quantum many-body approaches such as variational Monte Carlo, tensor-network approaches beyond DMRG, or the recently developed pseudo-Majorana functional renormalization group~\cite{Niggemann-2021}, which could shed light on the nature of the c$120^\circ$-QSL and QSL-VBC transitions from complementary numerical perspectives.

Furthermore, the problem of the microscopic characterization of the nature of the QSL can be addressed by a projective symmetry group classification~\cite{Wen-2002} of fermionic mean-field Ans\"atze with $U(1)$ and $\mathds{Z}_{2}$ low-energy gauge groups. These Ans\"atze could be analyzed within the pf-FRG framework itself by using the low-energy effective vertex functions (instead of the bare couplings) within a self-consistent Fock-like mean-field scheme to compute low-energy theories for emergent spinon excitations~\cite{Hering-2019,hering2022}. An alternate treatment would be to perform Gutzwiller projections yielding variational wave functions whose relative energetic competitiveness and spectrum of excitations could be assessed within a Monte Carlo framework~\cite{Iqbal-2013,Iqbal-2016_tri,Ferrari-2023}.

The numerical data shown in the figures are available on
Zenodo~\cite{data}.

{\it Acknowledgements.} 
Y.I. acknowledges helpful discussions with Pratyay Ghosh, Tobias M\"uller, Ronny Thomale, Rhine Samajdar, and Karlo Penc. The work of Y.I. was performed in part at the Aspen Center for Physics, which is supported by National Science Foundation grant PHY-2210452. The participation of Y.I. at the Aspen Center for Physics was supported by the Simons Foundation. The research of Y.I., C.H., and S.T. was supported, in part, by the National Science Foundation under Grant No. NSF PHY-1748958. Y.I. acknowledges support from the ICTP through the Associates Programme and from the Simons Foundation through grant number 284558FY19, IIT Madras through the Institute of Eminence (IoE) program for establishing QuCenDiEM (Project No. SP22231244CPETWOQCDHOC), the International Centre for Theoretical Sciences (ICTS), Bengaluru, India during a visit for participating in the program “Frustrated Metals and Insulators” (Code: ICTS/frumi2022/9). Y.I. acknowledges the use of the computing resources at HPCE, IIT Madras. 
L.G. thanks IIT Madras for funding a three-month stay through an IoE International Graduate Student Travel award, where this project was initiated and worked on in the early stages.
S.T. and L.G. acknowledge support from the Deutsche Forschungsgemeinschaft (DFG, German Research Foundation), within Project-ID 277146847, SFB1238 (projects C02 and C03). S.T. and L.G. acknowledge usage of the JUWELS cluster at the Forschungszentrum J\"ulich and the Noctua2 cluster at the Paderborn Center for Parallel Computing (P$\mathcal{C}^{2}$).

\bibliography{mll_nn} 

\end{document}